\pdfoutput=1
\documentclass[iop,apjl]{emulateapj}
\usepackage{apjfonts} 
\usepackage{times}
\usepackage{amssymb,amsmath}

\usepackage[breaklinks,colorlinks,urlcolor=blue,citecolor=blue,linkcolor=blue]{hyperref}
\usepackage[all]{hypcap}
\usepackage{graphicx} 
\usepackage{xspace}
\usepackage{paralist}
\usepackage{enumitem}

\newcommand{\ie}{{\it i.e.}\xspace}
\newcommand{\eg}{{\it e.g.}\xspace}

\newcommand{\etal}{{\it et al.}\xspace}
\newcommand{\kpc}{\ensuremath{\,{\rm kpc}}\xspace}

\newcommand{\au}{\ensuremath{\,{\rm a.u.}}\xspace}
\newcommand{\Gyr}{\ensuremath{\,{\rm Gyr}}\xspace}
\newcommand{\kms}{\ensuremath{\,{\rm km}\,{\rm s}^{-1}}\xspace}

\newcommand{\days}{\ensuremath{\,{\rm days}}\xspace}
\renewcommand{\arcmin}{\ensuremath{\,{\rm arcmin}}\xspace}

\newcommand{\Iband}{$I$-band\xspace}

\newcommand{\msun}{\ensuremath{{M_\odot}}\xspace}

\newcommand{\nevents}{{\ensuremath{2861}\xspace}}
\newcommand{\longtelim}{\ensuremath{200\,{\rm days}\xspace}}
\newcommand{\shorttelim}{{2\,{\rm days}\xspace}}

\newcommand{\ml}{\ensuremath{M_l}\xspace}

\newcommand{\ds}{\ensuremath{D_s}\xspace}
\newcommand{\dl}{\ensuremath{D_l}\xspace}

\newcommand{\te}{\ensuremath{t_E}\xspace}
\newcommand{\re}{\ensuremath{R_E}\xspace}
\newcommand{\ams}{\ensuremath{\alpha_{\rm{ms}}}\xspace}
\newcommand{\abd}{\ensuremath{\alpha_{\rm{bd}}}\xspace}
\newcommand{\Is}{\ensuremath{I_s}\xspace}
\newcommand{\Pmodel}{P17\xspace}

\newcommand{\amsfinal}{\ensuremath{\ams=1.31\pm0.10|_{\rm stat}\pm0.10|_{\rm sys}}\xspace}
\newcommand{\abdfinal}{\ensuremath{\abd=-0.7\pm0.9|_{\rm stat}\pm0.8|_{\rm sys}}\xspace}
\newcommand{\mcfinal}{\ensuremath{M_c=(0.17\pm0.02|_{\rm stat}\pm0.01|_{\rm sys})\msun}\xspace}
\newcommand{\sigmamfinal}{\ensuremath{\sigma_m=0.49\pm0.07|_{\rm stat}\pm0.06|_{\rm sys}}\xspace}

\shorttitle{The IMF Measured From OGLE-III Microlensing Timescales} 
\shortauthors{Wegg \etal}
\begin{document}
\title{The Initial Mass Function of the Inner Galaxy Measured From OGLE-III Microlensing Timescales}
 
\author{Christopher Wegg\altaffilmark{1}, Ortwin Gerhard\altaffilmark{1} and Matthieu Portail\altaffilmark{1}}
\affil{$^1$Max-Planck-Institut f\"ur Extraterrestrische Physik}
\affil{Giessenbachstrasse, 85748 Garching, Germany}
\email{wegg@mpe.mpg.de}


\begin{abstract}
We use the timescale distribution of $\sim 3000$ microlensing events measured by the OGLE-III survey, together with accurate new made-to-measure dynamical models of the Galactic bulge/bar region, to measure the IMF in the inner Milky Way. The timescale of each event depends on the mass of the lensing object, together with the relative distances and velocities of the lens and source. The dynamical model provides statistically these distances and velocities allowing us to constrain the lens mass function, and from this to infer the IMF. Parameterising the IMF as a broken power-law, we find slopes in the main sequence  \amsfinal and brown dwarf region \abdfinal where we use a fiducial $50\%$ binary fraction, and the systematic uncertainty covers the range of binary fractions $0-100\%$. Similarly for a log-normal IMF we conclude \mcfinal and \sigmamfinal. These values are very similar to a Kroupa or Chabrier IMF respectively, showing that the IMF in the bulge is indistinguishable from that measured locally,  despite the lenses lying in the inner Milky Way where the stars are mostly $\sim 10\Gyr$ old and formed on a fast $\alpha$-element enhanced timescale. This therefore constrains models of IMF variation that depend on the properties of the collapsing gas cloud.
\end{abstract}


\section{Introduction}

The present day mass function (PDMF), \ie the number of stars as a function of mass, is of importance in many areas of astronomy. For example it is the key ingredient in inferring the stellar masses of galaxies from their light. The luminosity is dominated by stars with masses close to the main sequence turnoff. The mass function is then necessary to infer the total stellar mass, including the more numerous fainter dwarfs and stellar remnants. The mass distribution of stars at birth, the initial mass function (IMF), is similarly important throughout astronomy, controlling not just the PDMF but also the return of gas to, and the enrichment of, the interstellar medium. Despite the importance of the IMF we have little  understanding of how it arises from the physics of the collapsing gas clouds. A variety of methods have been used to measure the PDMF and infer the IMF. The most direct measurements are from star clusters and field stars in the solar neighbourhood. Here the counting of stars  makes the measurement fairly robust outside the lowest mass brown dwarfs \citep[see \eg][for recent reviews]{Bastian:10,Krumholz:14,Offner:14}.

There are grounds for suspecting IMF variation with redshift or formation timescale because of dependence on \eg the temperature and density of the collapsing interstellar gas \cite[\eg][]{Bastian:10}. There has therefore been a great effort to extend our knowledge of the IMF outside the solar neighbourhood. Mass dependent absorption features in extremely high signal to noise spectra have been used to inter the PDMF, suggesting that it may vary in massive ellipticals \citep[\eg][]{vanDokkum:12,Conroy:12}. Other complimentary methods to probe the IMF in external galaxies estimate dynamical masses and break the degeneracy with dark matter either through lensing \citep{Dutton:12} or population expectations \citep{Cappellari:12,Thomas:11}.

Bulge microlensing is a unique tool for measuring the mass function in the inner Milky Way (MW), where the majority of stars formed quickly at redshift $z>1$ \citep[\ie they are enhanced in $\alpha$ elements and mostly $\sim10\Gyr$ old,][]{Rich:13review}. Microlensing events occur when a background source star passes in projection within the Einstein radius of a nearer star or stellar remnant and the light from the background star is therefore amplified. The level of amplification is purely geometrical and tells us no useful Galactic information. The timescale of each event however depends on several factors: the relative proper motion of the lens and source star, their distances, and the mass of the lens. Normally there is insufficient information in each event to infer all these, and so the lens mass, and the distances and velocities of the lens and source are degenerate. Dynamical models are however able to statistically provide the expected distances and velocities of microlensing events, and therefore from the distribution of microlensing timescales the lens mass distribution can be measured.

This method has been used to infer lens mass distributions several times in the literature. Both by taking moments of the distributions \citep{Jetzer:94,Grenacher:99}, and by fitting the timescale distribution directly from a galactic model together with a parametric IMF \citep{Han:96,Zhao:96,Bissantz:04,CalchiNovati:08,Moniez:17}. This work follows the latter approach. 
 
Very recently the two most important ingredients in determining the IMF from the microlensing timescale distribution have been greatly improved, which motivates this work. Firstly \citet[][hereafter \Pmodel]{Portail:17} presented the first dynamical model fitted to extensive photometric and kinematic data across the bulge, bar and inner disk of the MW. This model represents a significant improvement over previously available models. Secondly \citet{Wyrzykowski:15} presented a uniform sample of 3718 microlensing events detected in Optical Gravitational Lensing Experiment (OGLE) III data, together with the necessary efficiency as a function of timescale. This is large increase over the samples of $\lesssim50$ events used in previous microlensing IMF measurements.



\section{Dynamical Model}
\label{sec:models}

We use the dynamical model derived from inner Galaxy data by \Pmodel using the made-to-measure method \citep{Syer:1996ts,DeLorenzi:07}. In this work an initially barred $N$-body model was fitted to a wide range of data consisting of: the 3D shape of the bulge measured by \citet{Wegg:13}, combined near-infrared star counts from the VVV, UKIDSS and 2MASS surveys \citep{Wegg:15}, and kinematics from the BRAVA \citep{Kunder:12} and ARGOS \citep{Ness:13IV} surveys. The resultant model was shown to be consistent with OGLE-II proper motion data \citep{Sumi:04}, important for this work which depends on these transverse velocity distributions. The reader interested in the full details of the dynamical model is referred to \Pmodel.

\begin{figure}
\includegraphics[width=0.95\linewidth]{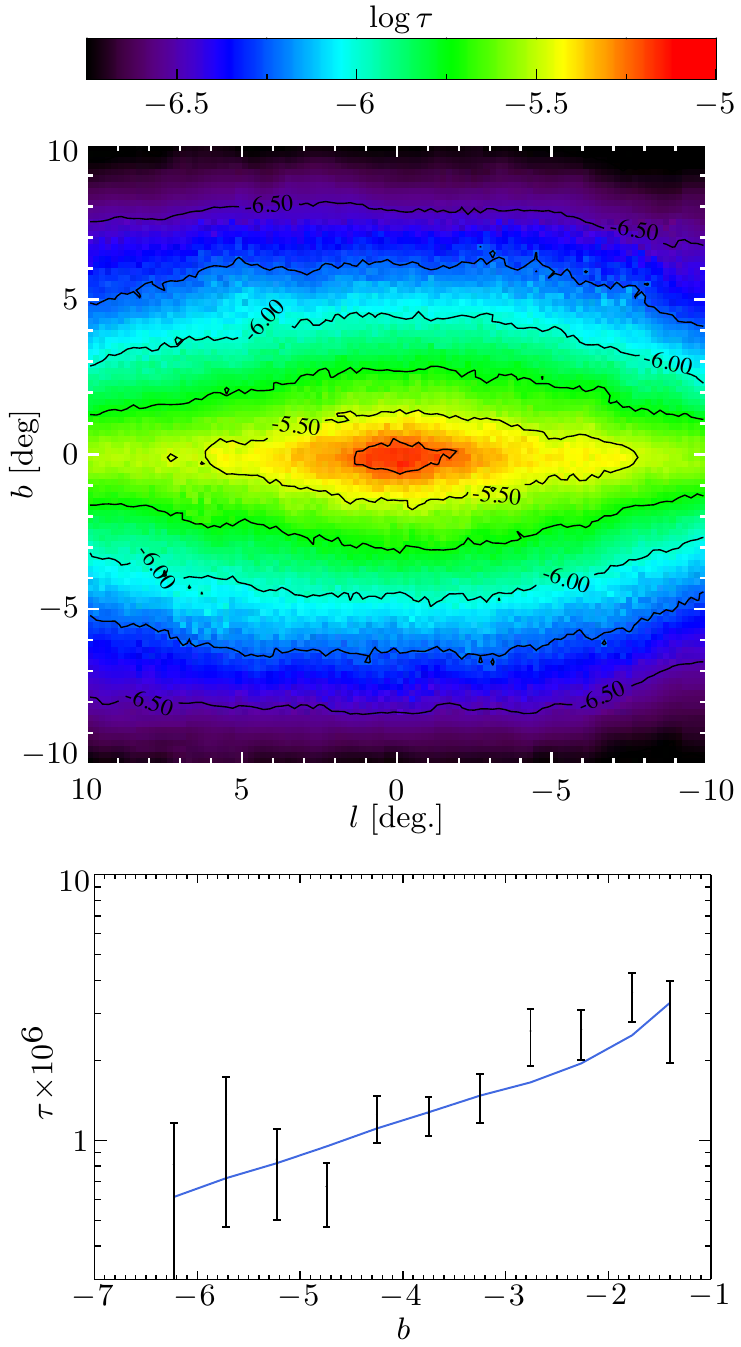}
\caption{The optical depth to microlensing of the dynamical model used in this work. Upper panel: Optical depth averaged over stars with unextincted source magnitude $14<I_s<19$. Lower panel: Comparison with the optical depths measured by \citet{Sumi:16}.  Both panels use the methods outlined in \citet{Wegg:16} with the updated dynamical model of \Pmodel. \label{fig:dynmod}}
\end{figure}

To provide confidence that the model reproduces the number of microlensing events we compare it to the MOA-II optical depth \citep{Sumi:16} as a function of Galactic latitude in \autoref{fig:dynmod}; the model was not fitted to this data and has a reduced $\chi^2$ of 0.9. We also show in \autoref{fig:dynmod} for reference how the optical depth to microlensing changes across the bulge in this model. The variation in our results from other P17 models (the boundary models), are significantly smaller than the systematic error from binaries (\autoref{sec:binaries}).
 
\section{Predicted Timescale Distribution}
\label{sec:tetheory}

The timescale \te of a microlensing event is the time for the source and lens in projection to cross the Einstein radius, \re, of a lens with mass \ml:
\begin{equation}
	\te=\re/V=\frac{1}{V}\sqrt{\frac{4G\ml\dl^2}{c^2}\left(\frac{1}{\dl}-\frac{1}{\ds}\right)}\label{eq:te}
\end{equation}
where \dl and \ds are the distances to the lens and source, and $V$ is the transverse velocity of the lens relative to the line-of-sight from Earth towards the source star.  


We define $\Gamma(\log\te|l,b,\Is,t_0)$ to be the probability distribution of $\log\te$ at given position $(l,b)$, and \Iband source magnitude \Is. The time of the event, $t_0$, is used to transform to the geocentric frame in which $V$ is defined.

We assume that the lens mass distribution is constant in space, and check this assumption in  \autoref{sec:results}. Because $\te\propto\sqrt{\ml}$, the timescale distribution is then given by the convolution \citep{Han:96,Wegg:16}
\begin{align}
\Gamma(\log\te|l,b,\Is,t_0)=\int&\gamma(\log\te-\frac{1}{2}\log\ml|l,b,\Is,t_0)\nonumber\\ 
&\quad\Phi(\log\ml)\sqrt{\ml}\,d\log\ml\label{eq:teconv}
\end{align}
where $\gamma(\log\te|l,b,\Is,t_0)$ is the timescale distribution predicted by the dynamical model if all lenses were $1\msun$, and $\Phi(\log M)$ the PDMF. 

To compute $\gamma(\log\te|l,b,\Is,t_0)$ we follow Section 3.2 of \cite{Wegg:16}. In brief, at each position $(l,b)$ we select the $10^4$ nearest $N$-body particles as potential lens-source pairs. 
For each lens-source pair $(j,i)$ we compute, using \autoref{eq:te}, the timescale $t_{E,ij}$ if the lens had mass $1\msun$. This timescale must then be correctly weighted using the $N$-body model.
To do so we use the optical depth for source particle $i$ to lensing by particle $j$:
\begin{equation}
	\tau_{ij}=\frac{4\pi G}{c^2\omega}M_j\left(\frac{1}{D_j}-\frac{1}{D_i}\right)\label{eq:taumc}
\end{equation}
where $\omega$ is the solid angle encompassed by the selected particles. This is the instantaneous probability that $j$ microlenses $i$. 
 $\tau_{ij}M_i/t_{E,ij}$\footnote{This corrects a missed $j$ index from $\tau_{ij}$ in Eq (16) of \citet{Wegg:16}.} is therefore proportional to the expected rate of events from pair $(i,j)$. All $(i,j)$ pairs are thus binned with weight $\tau_{ij}M_i/t_{E,ij}$ as a function of $\log\,t_{E,ij}$ and source distance modulus $\mu_{s,i}$. Each $\mu_{s,i}$-column of the resultant matrix corresponds to the timescale distribution for that source distance modulus. However only source magnitudes are known, and not distances. We therefore convolve this with the \Iband luminosity function of a 10\Gyr old population to provide the rate of events at this position and time as function of source magnitude and timescale: $\gamma(\log\te|l,b,t_0,\Is)$. Finally because here we are only interested in the distribution of timescales we normalise so that $\int\gamma(\log\te|l,b,\Is,t_0)\,d\log\te=1$.
\begin{figure}
\includegraphics[width=\linewidth]{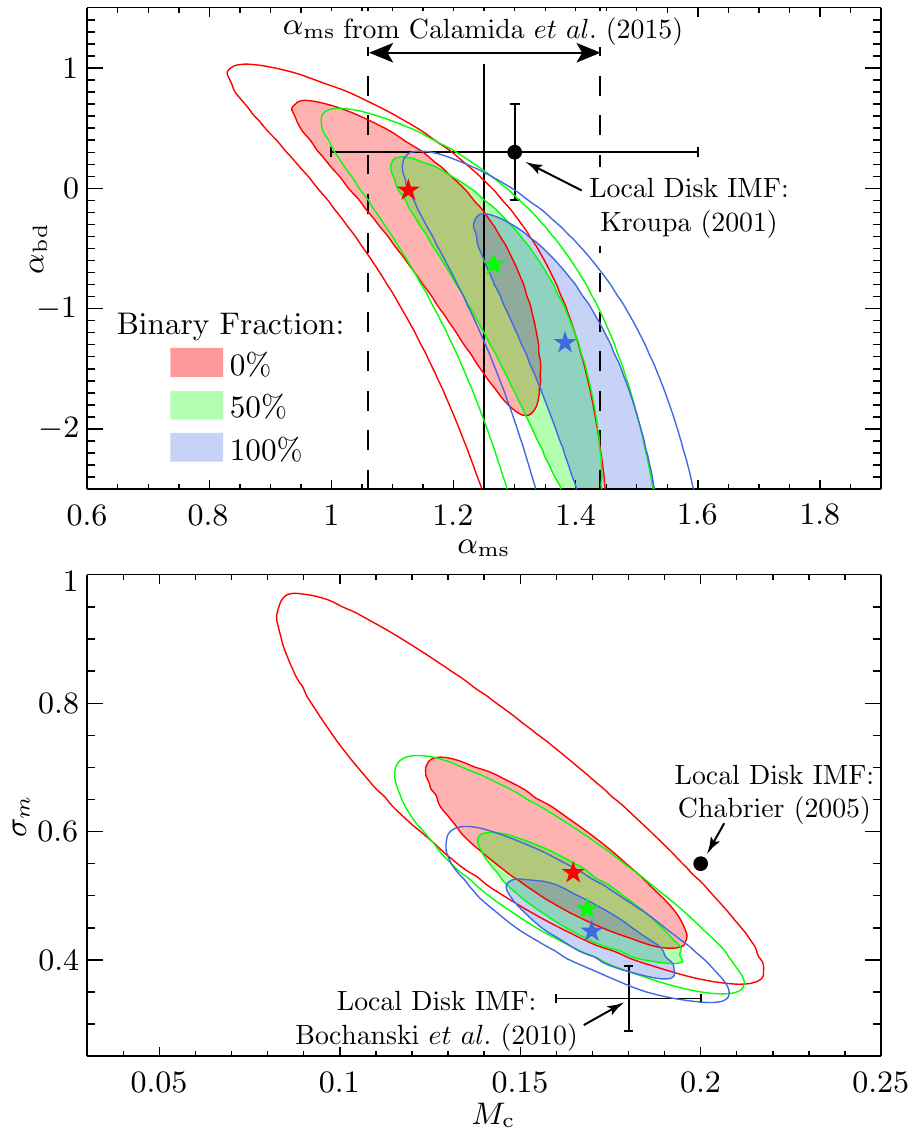}
\caption{Upper panel: $1\sigma$ and $2\sigma$ contours of the main sequence, \ams, and brown dwarf, \abd, slopes of \autoref{eq:kroupaimf}. Red corresponds to 0\% binary fraction, green to 50\%, and blue 100\%. The position of the maximum likelihood is shown as a star and the \citet{Kroupa:01} IMF as a black point \citep[errors from][]{Kroupa:13}. The $\ams=-1.25\pm0.19$ measured by \citet{Calamida:15} in the bulge is shown as the vertical lines. Lower panel: Similar for log-normal IMF. The values from \citet{Chabrier:05} and \citet{Bochanski:10} are shown in black. \label{fig:telike}}
\end{figure}

\section{Fitting the Mass Function}
\label{sec:mffit}

We utilise the timescales of the events provided by \citet{Wyrzykowski:15}. We remove highly blended events with blending proportion $f_s<0.2$ because, as discussed there, these display a bias towards longer timescales; this leaves~\nevents~events. When fitting we maximise the likelihood 
\begin{equation}
\log\mathcal{L}=\sum_{{\rm events}~i}\log\mathcal{L}_i~.
\end{equation}
The likelihood of an individual event with timescale $t_{E,i}\pm\sigma_{t_E,i}$ at $(l_i,b_i)$ with magnitude $I_{s,i}$ is 
\begin{align}
\mathcal{L}_i=\frac{1}{\sigma_{t_E,i}\sqrt{2\pi}}\int&\exp\left[-\frac{(t_{E,i}-\te)^2}{\sigma_{\te,i}}\right]\Gamma(\log\te|l_i,b_i,I_{s,i},t_{0,i})\nonumber\\
&  \times\mathcal{E}(\log\te)\,d\log\te
\end{align}
where  $\mathcal{E}(\te)$ is the detection efficiency of events with timescale \te  \citep[also provided by][]{Wyrzykowski:15}, and $\Gamma(\log\te|l_i,b_i,{\Is}_i,t_{0,i})$ is the predicted timescale distribution computed as described in the previous section. We do not focus on the shortest timescale events here, which are possibly produced by planet size mass lenses \citep{Sumi:11,Clanton:17}. We also do not fit the longest timescale events; the point lens-point source model fitted in \citet{Wyrzykowski:15} does not take account of parallax as the earth moves around the sun. These are instead the focus of dedicated searches \citep[\eg][]{Wyrzykowski:16}. We therefore consider only $\shorttelim<\te<\longtelim$.

We consider IMFs of broken power-law form
\begin{align}
dN&=\Phi(\log M)\,d\log M\nonumber\\
&\propto M^{-\alpha}\,dM\mbox{~~where~~}\label{eq:kroupaimf}\\
\alpha&=\abd\mbox{~for~}0.01M_\odot\leq M<0.08M_\odot\nonumber\\
\alpha&=\ams\mbox{~for~}0.08M_\odot\leq M<0.5M_\odot=M_{\rm break}\nonumber\\
\alpha&=2.3\mbox{~for~}0.5M_\odot\leq M<100M_\odot~~.\nonumber 
\end{align}
A \citet{Kroupa:01} IMF corresponds to $\ams=1.3$ and $\abd=0.3$. In the bulge  $\ams=1.25\pm0.19$ was measured from star counts with an only slightly different break ($M_{\rm break}=0.56M_\odot$, \citealt{Calamida:15}), quite close to the $\ams=1.43\pm0.13$ measured in the bulge at higher latitude by \citet{Zoccali:00}. 

We also consider IMFs of log-normal form
\begin{align}
\Phi(\log M)&\propto\exp\left\{\frac{-(\log M-\log M_c)^2}{2\sigma_m^2}\right\}\mbox{~for~}M<1.0M_\odot\label{eq:chabrierimf}\\
\alpha&=2.3\mbox{~for~}1.0 M_\odot\leq M<100M_\odot~~.\nonumber 
\end{align}
The \citet{Chabrier:05} IMF has $M_c=0.2\msun$ and $\sigma_m=0.55$. 

We transform from IMF to PDMF using a $10\Gyr$ population and the remnant prescription of \citet{Maraston:98}. We have checked that our conclusions are insensitive to these choices by comparing to an exponentially declining star formation rate and the remnant prescription of \citet{Percival:09}.

\begin{figure}
\includegraphics[width=\linewidth]{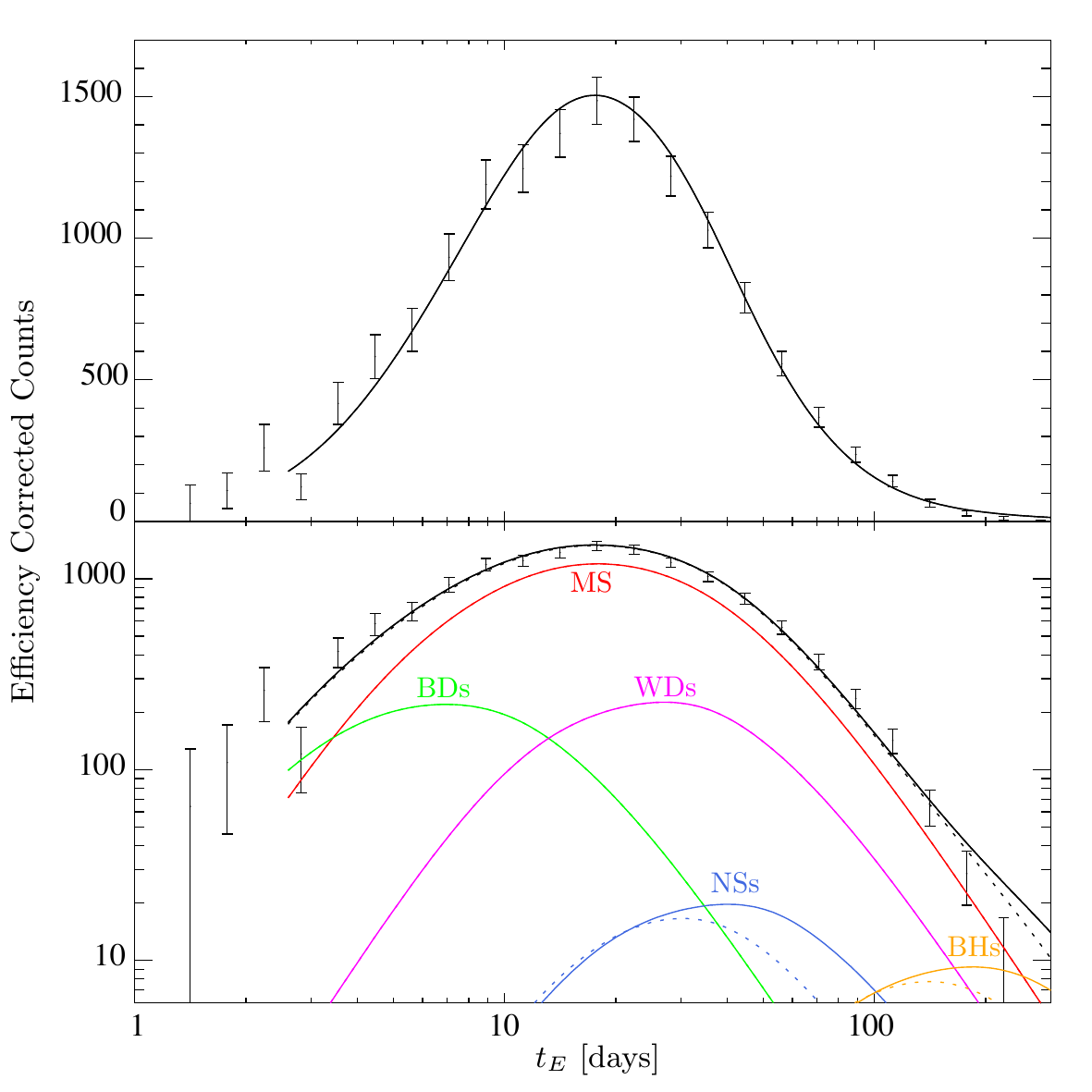}
\caption{The efficiency-corrected timescale distribution from \citet{Wyrzykowski:15} compared to the best fitting power-law IMF with 0\% binary fraction. The lower panel shows the contribution to the model distribution from brown dwarfs (green), main sequence stars (red), white dwarfs (magenta), neutron stars (blue) and black holes (orange). The dotted lines show the same model but giving the neutron stars and black holes a natal kick of dispersion $\sigma_{\rm{1d}}=190\kms$.\label{fig:tedist}}
\end{figure}

In both \autoref{eq:kroupaimf} and \autoref{eq:chabrierimf} we enforce continuity of the mass function at the breaks (but not its derivative) leaving two free parameters, either $(\abd,\ams)$ or $(\log M_c,\sigma_m)$. Throughout we have assumed flat priors on these quantities. In \autoref{fig:telike} we show the resultant likelihood contours in red of $(\abd,\ams)$ above and $(\log M_c,\sigma_m)$ below. The best fitting timescale distribution is compared to the data in \autoref{fig:tedist}.

\subsection{The impact of binaries}
\label{sec:binaries}

Often when measuring the IMF corrections must be made for the effect of unresolved binaries. The typical Einstein radius of microlensing events towards the bulge is $2\au$. Binaries with this separation are often poorly fit by the single lens model, frequently presenting caustic crossings. These have been removed from the sample of \citet{Wyrzykowski:15} where only events which were well fit by the single lens model are included. Binaries with separation significantly wider than \re will be well resolved, however binaries with smaller separation will be unresolved. Here we assess the impact of this on the IMF. To do so we perform a binary population synthesis using the code of \citet{Hurley:02}. 

We evolve a population of solar metallicity binaries characterised by primary mass $M_1$, secondary mass $M_2$, separation $a$, eccentricity $e$ for 10\Gyr. From solar neighbourhood data \citep{Raghavan:10} we assume:\begin{inparaenum}[(a)]
 	\item The period distribution $P(\log P)\propto\exp\left[\left(\log P-\log P_0\right)^2/2\sigma_P^2\right]$ with $\log(P_0/{\rm days})=5.0$ and $\sigma_P=2.3$.
 	\item A flat mass ratio, $P(q\equiv M_1/M_2)=\mbox{const.}$, between 0 and 1.
 	\item An initially thermal eccentricity distribution, $P(e)\propto e$ between 0 and 1.
 	\item The primary mass distribution $f(M_1)$ so that the distribution of individual stellar masses matches the single stellar case. Inclusion of the distribution of secondaries with their mass distribution means this differs from choosing $f(M_1)$ to be \autoref{eq:kroupaimf} or \autoref{eq:chabrierimf}.
 	\item  Finally because close brown dwarf companions of solar type stars are extremely rare (the brown dwarf desert, \citealt{Grether:06}) we remove secondaries of mass $M<M_{\rm BD}=0.08\msun$.
\end{inparaenum}

We then treat binaries separated by $<3.7\sqrt{(M_1+M_2)/\msun}\au$ after 10\Gyr of evolution as unresolved (where the lens is the total system) and wider binaries as resolved (where the lens is one component of the system). Since we are concerned with estimating the size of the correction more sophisticated simulations are unwarranted.

We show in \autoref{fig:telike} the effect of changing the IMF to binary fractions of 50\% and 100\% (50\% corresponding to two-thirds of all stars born in binaries). 
We use the locally measured 50\% as our fiducial binary fraction \citep{Raghavan:10}, but because the bulk of the lens population are M-dwarfs and the binary fraction of such low-mass stars in the bulge has not been measured, we also consider 0\% and 100\% binary fractions. 
\subsection{Summary of Results}
\label{sec:results}

Computing confidence intervals from the likelihood assuming flat priors on the IMF parameters gives the values in \autoref{tab:parms}. Different binary fractions and IMF forms lie within $\Delta\ln\mathcal{L}<1.2$ of each other and therefore given the uncertainties in binary population we do not significantly prefer any.

\autoref{fig:telike} shows that formally the IMF differs statistically at the $\sim2\sigma$ level with the local disk values from \citet{Kroupa:01} and \citet{Chabrier:05}. This suggests that the average mass is lower than those fiducial values. However the differences of $\Delta\ams\sim0.1$ and $\Delta M_c\sim0.02\msun$ are within the error budget of local IMF determinations. We therefore conclude that the inner MW IMF measured here is indistinguishable from that measured in the local disk.

Taking the locally measured 50\% binary fraction as the fiducial value we conclude for the power-law IMF \amsfinal and \abdfinal where the systematic uncertainty covers the range of binary fractions $0-100\%$. Similarly for the log-normal IMF we conclude \mcfinal and \sigmamfinal. 

Note that from \autoref{fig:telike} $\abd<1$ at $1\sigma$ for all binary fractions meaning that the number (and the mass) per logarithmic mass interval is falling towards lower mass.

\begin{table}
\begin{center}
	\caption{Parameters and their symmetric $1\sigma$ errors for the power-law (\autoref{eq:kroupaimf}) and log-normal IMF (\autoref{eq:chabrierimf}) as a function of binary fraction. $\Delta\ln\mathcal{L}$ is the maximum log-likelihood compared to the model with the highest log-likelihood: the log-normal IMF with 100\% binary fraction.\label{tab:parms}}
\begin{tabular}{cccc}
\tableline
Binary Fraction&$\alpha_{\rm ms}$&$\alpha_{\rm bd}$&$\Delta\ln\mathcal{L}$\\
\tableline
0\%&$1.21\pm0.12$&$-0.02\pm0.91$&1.1\\
50\%&$1.31\pm0.10$&$-0.65\pm0.89$&0.5\\
100\%&$1.39\pm0.09$&$-1.45\pm0.76$&0.4\\
\tableline
Binary Fraction&$M_c$&$\sigma_m$&$\Delta\ln\mathcal{L}$\\
\tableline
0\%&$0.162\pm0.025$&$0.54\pm0.10$&1.2\\
50\%&$0.166\pm0.018$&$0.49\pm0.07$&0.5\\
100\%&$0.169\pm0.015$&$0.45\pm0.05$&0.0\\
\tableline	
\end{tabular}
\end{center}
\end{table}

\begin{figure*}
\includegraphics[width=\linewidth]{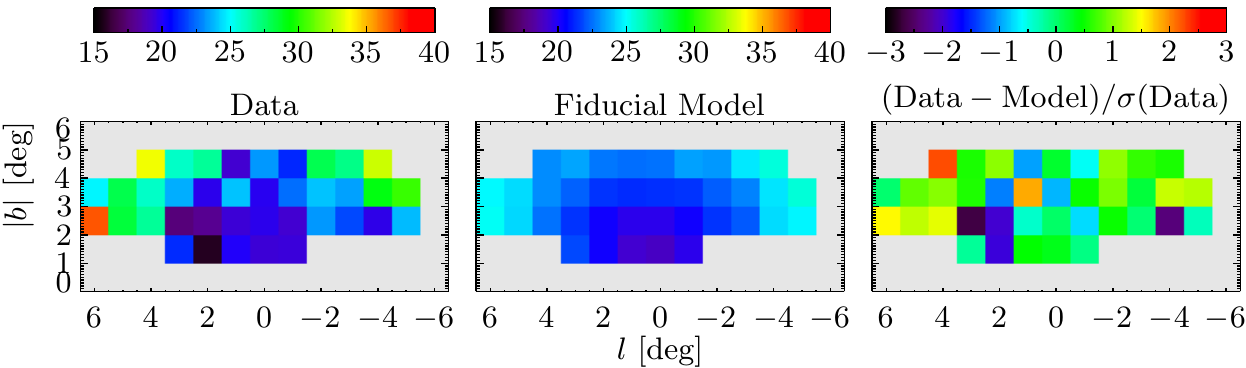}
\caption{Left panel: The mean log timescale in days measured by OGLE-III as a function of position. Central panel: The same quantity computed for the best fitting power-law IMF model. Right panel: The difference in number of sigma. \label{fig:tevspos}}
\end{figure*}

As a consistency check \autoref{fig:tevspos} shows the variation of $\left<\log\te\right>$ in Galactic coordinate bins. The trend seen in the data of shorter timescales close to the Galactic centre is reproduced by the model. Significant changes in the PDMF over this area, or differences between dynamical model and the MW, would show as inconsistent model variation.

\subsection{Stellar Mass Black Holes}

We show in the lower panel of \autoref{fig:tedist} the timescale distribution in our fiducial model, and of each type of lens separately. While the overall fit is generally excellent, for the longest events with $\te\gtrsim100\,\mbox{days}$, the fiducial model predicts more events than observed. There are several possible explanations: 
\begin{enumerate}
\item Not including parallax motion when fitting the light curve in the simple lens model fitted by \citet{Wyrzykowski:15} could result in longer timescale events being inefficiently detected or \te misestimation.
\item This discrepancy occurs where black hole lenses become important and so could indicate the remnant prescription produces too much mass in BHs.
\item Part of the discrepancy could be because BHs and NSs are likely to receive significant natal kicks \citep[\eg][]{Hansen:97,Repetto:12}. The distribution of remnants will therefore expand in space, particularly in the disk, away from the low-latitude sight-lines towards the bulge. In addition the remnants that do microlens would have shorter timescales because of their larger velocities. We show the effect of a Maxwellian natal kick distribution of $\sigma_{\rm{1d}}=190\kms$ as the dotted lines in the lower panel of \autoref{fig:tedist}. The distribution and kinematics of NS and BHs were computed by applying an impulsive kick from this distribution to the stars in the model, and integrating forwards 1.8\Gyr. All other details of the best fitting model were kept fixed.
\end{enumerate}
The other dynamical models tested from P17 give very similar long timescale behaviour \cite[as asymptotically expected,][]{Mao:96} and so the discrepancy is unlikely to result from the choice of model. We have also checked that reducing the upper timescale limit to $100\days$ does not change our conclusions, altering the best fitting model by well within the statistical errors.

\section{Discussion}
\label{sec:discuss}

\subsection{Comparison to other methods of IMF determination}

Here we measure the IMF towards the inner MW using microlensing. The most directly comparable method is that of star counts towards the bulge. Microlensing has the advantage that it readily reaches lower masses than accessible to star counts, even down to brown dwarf and planet masses.
 
Local measurements using star counts in the field and nearby star clusters have uncertainties due to unresolved stellar multiplicity and, particularly for brown dwarfs, masses that can depend significantly on uncertain evolutionary models.

In comparison microlensing timescales are fundamentally sensitive directly to the mass of the lens. Conversion of timescale to mass does require a Galactic model, however because of advances in modelling techniques and the wealth of kinematic data on the inner Galaxy  this uncertainty is now small. In addition only binaries with separation less than a few \au are unresolved, making inferences on the mass of the individual stars more straightforward compared to photometric IMF determinations. 

\subsection{The Mass-to-Light and Mass-to-Clump Ratios of the Bulge}

Even in deep star counts, only stars with $M\gtrsim0.15\msun$ are presently observable in the Bulge and therefore when calculating mass-to-light assumptions must be made about the contribution of lower mass stars and remnants. Microlensing is sensitive to these unseen objects and so the measurements here can be combined with star counts to robustly measure the mass-to-light. \citet{Calamida:15} measured the mass of visible stars with $>0.16\msun$ in the SWEEPS field to be $137,500\pm23,400\msun$. These stars produced $L_{F814W}=104,000\pm2,000L_\odot$ and $L_{F606W}=71,000\pm1,400L_\odot$. Using the PDMF from the power-law IMFs consistent with the microlensing data gives $(M/L)_{F814W}=(2.2\pm0.3)\msun/L_\odot$ and $(M/L)_{F606W}=(3.2\pm0.6)\msun/L_\odot$ where the uncertainties are dominated by the uncertainty in the mass of the visible stars.

The dynamical models of the bulge in \Pmodel used red clump stars (RCGs) as tracers of the MW's structure and therefore used a mass-to-clump ratio analogous to mass-to-light in external galaxies. \Pmodel used $(1000\pm100)\msun/{\rm{RCG}}$ but this required assumptions about the mass of low mass stars and remnants. It was based on  $2255\msun/\arcmin^2$ for stars $>0.15\msun$ at $(l,b)=(0\deg,-6\deg)$ and $4.0\pm0.4\mbox{RCGs}/\arcmin^2$ in the same direction \citep{Zoccali:00}. Computing the mass-to-clump similarly to the mass-to-light above gives $(960\pm100)\msun/{\rm{RCG}}$. Because this agrees with the mass-to-clump used by \Pmodel it strengthens the argument made there for a low dark matter fraction in the Bulge (\citealt{Wegg:16}; \Pmodel) which to be reconciled with the circular velocity and dark matter estimates locally requires a core in the MW's dark matter halo	. 

\subsection{The Inner MW IMF in Context}

The \Pmodel model predicts that the lenses have mean distance 6.3\kpc, with a mean galactocentric radius of 2.0\kpc. As such the PDMF measured from the microlensing timescale distribution probes the IMF in the bulge and disk of the inner Galaxy. It does so directly through the lens mass, uniquely down to low masses. The much larger sample of microlensing events from OGLE-III, combined here with the newly constructed dynamical models of the inner Galaxy, provide more stringent constraints on the IMF of the inner Galaxy than previously possible.

We find that the IMF in the inner Galaxy is consistent with those measured locally by \citet{Kroupa:01} and \citet{Chabrier:05}. However the inner Galaxy formed on a much shorter timescale than the local disk: it is $\alpha$-enhanced with a formation timescale $\sim0.5\Gyr$ \citep{Matteucci:14}. It is also significantly older: most bulge stars are $\sim10\Gyr$ \citep{Clarkson:11,Bensby:17}.  The consistency of the IMF between the inner MW measured here and the local disk  therefore places stringent constraints on star formation models where the IMF varies according to the properties of the parent molecular gas cloud \citep[see \eg][and references therein]{Guszejnov:17}.

\acknowledgments
The authors thank the anonymous referee for their timely and thorough reading of the manuscript.



\label{lastpage}
\end{document}